\documentclass[a4paper]{article}
\usepackage{graphicx}
\usepackage[margin=0.6in]{geometry}
\usepackage{amsmath}
\usepackage{amssymb}
\usepackage{cite}

\usepackage{multicol} 
\usepackage{caption}
\usepackage{hyperref}
\newcommand{\footremember}[2]{%
    \footnote{#2}
    \newcounter{#1}
    \setcounter{#1}{\value{footnote}}%
}
\newcommand{\footrecall}[1]{%
    \footnotemark[\value{#1}]%
} 

\title{It sounds like an action potential: unification of electrical, chemical and mechanical aspects of acoustic pulses in lipids}

\author{Matan Mussel\footremember{TU-Dortmund}{Department of Physics, Technical University of Dortmund, 44227 Dortmund, Germany}\footremember{AugsburgUni}{Department of Physics, University of Augsburg, 86159 Augsburg, Germany}\footnote{Correspondence to: matan.mussel@tu-dortmund.de}%
  \and Matthias F.~Schneider\footrecall{TU-Dortmund} %
  }

\begin{document}

\maketitle

\begin{abstract}
In an ongoing debate on the physical nature of the action potential, one group adheres to the electrical model of Hodgkin and Huxley, while the other describes the action potential as a non-linear acoustic pulse propagating within an interface near a transition. However, despite remarkable similarities, acoustics remains a non-intuitive mechanism for action potentials for the following reason. While acoustic pulses are typically associated with the propagation of density, pressure and temperature variation, action potentials are most commonly measured electrically. Here, we show that this discrepancy is lifted upon considering the electrical and chemical aspects of the interface, in addition to its mechanical properties. Specifically, we demonstrate how electrical and pH aspects of acoustic pulses emerge from an idealized description of the lipid interface, which is based on classical physical principles and contains no fit parameters. The pulses that emerge from the model show striking similarities to action potentials not only in qualitative shape and scales (time, velocity and voltage), but also demonstrate saturation of amplitude and annihilation upon collision.
\end{abstract}

{\bf Keywords:} Action potential, lipid interface, acoustics, phase transition, transmembrane potential \hfill \break

%

\begin{multicols}{2}
\section{Introduction}
Since Luigi Galvani's discovery in 1791 of muscle twitching upon electric excitation\cite{galvani1791d}, the phenomenon now known as {\it action potential} has been demonstrated to play a crucial functional role in the behavior of many living organisms. Action potentials (APs) are pulses that propagate along the cell surface and cause a transient change that can be captured by many observables. The most common observation is a characteristic change in the electric potential difference across the cell surface. But, in addition, the pulse propagates a mechanical deformation, a change in the intrinsic optical properties, as well as a local heating followed by cooling\cite{Tasaki1999}. 
 
Although the mechanism of APs has been investigated through various approaches, an active debate is still ongoing\cite{Tasaki1999, Kaufmann1989, Heimburg2005, Mussel2018}. In this paper we are concerned with the conjecture that APs are acoustic pulses that propagate along the lipid bilayer at the cell surface, and cause a transient local phase transition\cite{Kaufmann1989, Heimburg2005}. Indeed, acoustic pulses have been demonstrated to propagate within lipid interfaces\cite{Griesbauer2009}, and such interfaces may exist in several liquid or crystalline phases separated by first or second order transitions\cite{Albrecht1978}. Specifically, acoustic pulses that traverse the melting transition between the so called {\it liquid-expanded} and {\it liquid-condensed} phase display unusual non-linear properties, that share many similarities with APs; including the time and velocity scales, qualitative pulse shape, a sigmoidal (``all-or-none'') response to excitation amplitude, as well as annihilation upon collision\cite{Shrivastava2014, Shrivastava2015, Shrivastava2017, Mussel2018}. 

\begin{figure*}[htb]
\centering
\includegraphics[width=0.7\linewidth]{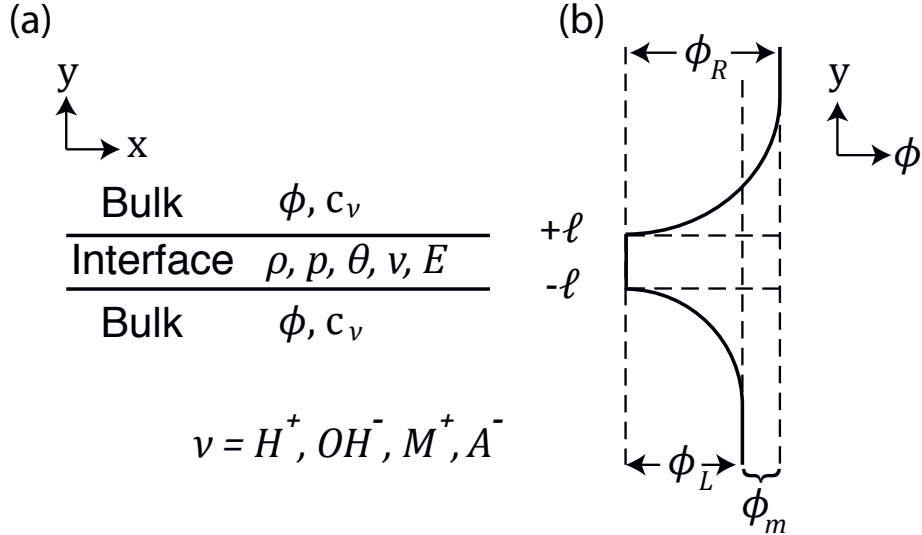}
 \caption{(a) The system geometry is a $2\ell$ wide flat interface, that separates two regions of bulk fluid. The interface is represented by its density $\rho$, pressure $p$, temperature $\theta$, velocity field $v$, and specific total energy $E$. The bulk contains four ionic species whose concentration is represented by $c_\nu$ (with $\nu=H^+, OH^-, M^+, A^-$) as well as the electric potential field $\phi$. (b) Sketch of the electric potential along the y-axis for a negatively charged interface with zero permeability to the mobile particles, and negligible electric field within the interface.}
 \label{Fig-01}
 \end{figure*}

The classical theory of acoustics portrays the longitudinal variation of the temperature, pressure and density of the system\cite{courant1948supersonic}. In lipids, however, the propagating state change is often strongly modified by electrical and chemical interactions with neighboring molecules. Particularly, ionizable groups on the head of the lipid molecules result in a net surface charge that tends to expand the membrane structure\cite{trauble1977membrane} and, in addition, generates electrical changes that are inseparable from the acoustic pulses\cite{Griesbauer2012a}. On the other hand, substances that adsorb on the acidic polar head; for example, protons, reduce the electric repulsion between the lipid molecules\cite{Trauble1974}. Thus, chemical changes at the lipid interface also accompany acoustic pulses\cite{Fichtl2016, Fichtl2018}. An understanding of these electro-chemical effects appears crucial to the development of a realistic theory of acoustics in lipid interfaces. 

In this work we study an idealized example that merges the electro-chemical with the thermo-mechanical properties of the interface. The analysis of electric variations that accompany acoustic pulses in lipids was previously addressed by Heimburg and Jackson\cite{Heimburg2005, Heimburg2012}. Their work presented solitonic solutions that arise from a small amplitude approximation of the lipid hydrodynamics. While the soliton model presents many valid arguments and captures the velocity and voltage scales, it fails to describe several important properties of the pulses, including the time scale, the qualitative shape, the saturation of amplitude at sufficiently strong excitations, and the annihilation of pulses upon collision. Here, we lift several limiting assumptions (for example, the small amplitude analysis) and consider a more general system that includes additional features (e.g., the effect of pH on the lipid interface). Specifically, we consider a synthesis of three theoretical ingredients, namely fluid mechanics, Maxwell's theory of electrodynamics and a phenomenological (macroscopic) constitutive equation. We demonstrate that the propagation of electrical and pH changes are inseparable from the density, pressure, and temperature aspects of longitudinal pulses in lipids. In addition, we show that the electrical aspect is comparable to APs in multiple features, including (1) the qualitative shape, (2) the time, velocity and voltage scales, (3) the sigmoidal response to excitation, and (4) annihilation upon collision. 

\section{Model Description}
The system under consideration is a lipid interface that separates two regions of bulk fluid. The interface is composed of acidic lipid molecules with a net negative charge in the ionized state. Acoustic pulses, that propagate along the interface, transiently modify the surface charge density, and consequently alter the electric fields that penetrate the bulk. These electric fields disturb mobile charged ions (protons included) that are dissolved in the bulk. In turn, the ions can associate with the lipid head and modify the state of the interface as well as the electric fields. Therefore, we hypothesize that acoustic pulses in lipids are accompanied by electric and chemical aspects. Specifically, we focus in this work on two common observables, the electric potential difference across the interface and the pH at the surface. 

\paragraph{Model equations.} This work extends a previous study on the thermo-mechanical aspects (i.e., temperature, density and pressure) of acoustic pulses within a medium near a phase transition\cite{Mussel2018}. The dynamics of the interface are governed by conservation laws; the conservation of mass, momentum and energy. The latter implies that the total entropy of the system is conserved\cite{courant1948supersonic}. We further adopt the assumption of continuity, and treat the system as a continuum of local thermodynamic states; i.e., each infinitesimal point (a fluid parcel) is associated with equilibrium thermodynamic quantities (e.g., temperature and pressure)\cite{prigogine1998modern}. An idealized geometry of a $2\ell$ wide flat interface was chosen for simplicity, and is depicted in figure \ref{Fig-01}a along with the model observables. The interface is characterized by five continuous variables, the specific volume ($w$), interfacial pressure ($p$), interfacial temperature ($\theta$), velocity ($v$), and specific total energy ($E$). The density of the interface is inversely proportional to the specific volume, $\rho=w^{-1}$. We use an idealized ansatz of these laws, following the work of Slemrod\cite{slemrod1984dynamic}.
\begin{equation}\label{eq:consLaws}
\begin{aligned}
\partial_t w&=\bar{w}\partial_h v, \\
\partial_t v&=\bar{w}\partial_h (\tau_1+\tau_2 ), \\
\partial_t E&=\bar{w}\left[\partial_h (\tau_1 v)+k\partial_h^2 \theta\right].
\end{aligned}
\end{equation}
The equations are given in the Lagrange frame, which encodes fluid parcels by the spatial coordinate $h$, and are later translated into the $x$ axis of Euler frame (laboratory frame) according to the cumulative mass of the fluid parcels\cite{courant1948supersonic}
\begin{equation}
x=\frac{1}{\bar{w}}\int w dh,
\end{equation}
with $\bar{w}$ a normalization factor that defines the relative scale of $h$ and $x$. In addition, $k$ is the coefficient of thermal conductivity, and the interfacial stresses,  $\tau_1$ and $\tau_2$, are
\begin{equation}
\begin{aligned}
\tau_1&=-p+\zeta\partial_h v, \\
\tau_2&=-C\partial_h^2 w,
\end{aligned}
\end{equation}
with $\zeta$ the dilatational viscosity and $C$ the capillarity coefficient, treated here as constant for simplicity. 
 
The assumption of continuity implies that phenomenological constitutive relations, obtained from static measurements, may be applied locally\cite{prigogine1998modern}. The constitutive relation of the melting transition is captured using a synthesis of two models, proposed by van der Waals and Tr{\"a}uble, respectively. 
\begin{equation}\label{eq:constitutiveEqs}
\begin{aligned}
p &= p_{vdW} + p_{tr}, \\
E &= E_{vdW} + E_{tr}, \\
\end{aligned}
\end{equation}
The van der Waals famous temperature-pressure-area relation qualitatively resembles the temperature dependent melting transition in lipids\cite{Albrecht1978}
\begin{equation}\label{eq:vdW}
\begin{aligned}
p_{vdW}&=\frac{k_B \theta}{m w - b} - \frac{a}{m^2 w^2},\\
E_{vdW}&=\frac{v^2}{2} + c_v \theta - \frac{a}{m^2 w} + \frac{C}{2}(\partial_hw)^2.
\end{aligned}
\end{equation}
Here, $k_B$ is the Boltzmann constant, $m$ is the mass of a fluid particle, $a$ is the average attraction between particles, $b$ is the volume excluded by a fluid particle, and $c_v$ is the specific heat capacity\cite{johnston2014thermodynamic}. The critical point of the phase transition can be expressed by three parameters: $m, a$ and $b$ (Eq.~(S12) in the Supplemental Materials).  
 
Tr{\"a}uble's model contains the additional electrostatic effect of the ionized polar heads of acidic lipids, that tend to expand the membrane structure\cite{trauble1977membrane}. The charge density at the surface, which arises from the ionized polar heads ($P^-$), may be reduced due to adsorption processes of charged mobile particles with the membrane (e.g., $H^+ + P^- \rightleftharpoons HP$). Therefore, the surface charge depends not only on the density of the lipid molecules, but also on the local concentration of protons and other particles that can bind to the membrane. Because the adsorption process time ($\lesssim10^{-8}$ s\cite{israelachvili2011intermolecular}) is much smaller than the time scale of acoustic pulses in lipids ($\gtrsim10^{-3}$ s\cite{Mussel2018}), we can safely assume that an adsorption equilibrium is attained at every moment of an acoustic pulse. An additional crucial assumption is that the two sides of the interface (hereafter {\it leaflets}) are asymmetric. In the Results section we shall demonstrate that the asymmetry of the interface plays an important role in the generation of a transmembrane potential difference. To capture the effect of adsorption on the surface charge density we use the Langmuir adsorption model
\begin{equation}
\sigma_\pm = -\frac{e}{mw}\frac{1}{1+\sum_\nu K_{\nu,\pm} c_{\nu,\pm\ell}}.
\end{equation}
The index $\pm$ denotes the respective leaflet; i.e., the charge density is given at $y=\pm\ell$. In addition, $K_{\nu,\pm}$ is the association constant of ion $\nu$ with the lipid molecules, and $c_{\nu,\pm\ell}=c_{\nu}(x,\pm\ell,t)$ is the local concentration of ion $\nu$ at the surface. Assuming that the two leaflets are independent implies a linear summation of their electrical contribution to the free energy, and as a result also to the interfacial pressure and energy
\begin{equation}\label{eq:constEqs}
\begin{aligned}
p_{tr} &= p_{tr,+} + p_{tr,-}, \\
E_{tr} &= w\left(p_{tr,+} + p_{tr,-}\right), \\
p_{tr,\pm} &= \frac{2k_B\theta }{e}\sigma_{D,\pm}\left[\sqrt{1+\left(\frac{\sigma_\pm}{\sigma_{D,\pm}}\right)^2} - 1\right].
\end{aligned}
\end{equation}
A detailed derivation of Tr{\"a}uble's model is given in the Supplemental Materials. The model parameter $\sigma_{D,\pm}$ is related to the Debye screening length ($\lambda_{D,\pm}$) according to 
\begin{equation}
\sigma_{D,\pm} = \frac{4\varepsilon_0\varepsilon_r^bk_B\theta}{e\lambda_{D,\pm}}, 
\end{equation}
with $\varepsilon_0$ the vacuum permittivity, $\varepsilon_r^b$ the relative permittivity of the bulk fluid, and $e$ is the elementary electric charge. The Debye length is 
\begin{equation}\label{eq:lambdaDpm}
\lambda_{D,\pm} = \left(\frac{\varepsilon_0\varepsilon_r^bk_B\theta}{e^2N_A\sum_\nu c_{\nu,\pm\infty}}\right)^{1/2},
\end{equation}
where $N_A$ is the Avogadro number, and $c_{\nu,\pm\infty}\equiv c_{\nu}(x,\pm\infty,t)$ is the ionic concentration sufficiently far from the interface. 

A physiological screening length ($\lambda_{D,\pm}$$\sim$1--10 nm) requires mobile charged particles at ionic concentration of $\sim$1--100 mM. However, at physiological pH ($\sim$7) protons alone are insufficient to account for the nano-scale screening length. Therefore, in addition to protons we consider a minimum set of three other ionic species; hydroxide ($OH^-$), one type of alkaline ion ($M^+$)  and one type of halogen ion ($A^-$). The latter two are given in physiological concentration ($\sim$1--100 mM) in the bulk fluid. For simplicity, only the protons can associate the interface in the present model. 

The model is completed by describing the dynamics of the additional components of the system, namely the charged mobile particles, the electromagnetic fields and the bulk fluid. To avoid seeing the forest for the trees, we neglect the fluid dynamics of the bulk and the magnetic field, since their contributions are not crucial to demonstrate the emergence of electrical and chemical aspects of acoustic pulses. We discuss their effect in the Discussion section, and stress that it should be taken into account in a more detailed future analysis. The temperature at the bulk was treated as a constant, $\theta_0$. 

The dynamics of the mobile particles are determined from the law of mass conservation. Following the Nernst-Planck model, ionic concentration gradients as well as electric fields are the main ingredients of ionic currents
\begin{equation}\label{eq:NernstPlanck}
\begin{aligned}
\partial_tc_\nu &= - \nabla \cdot \vec{j}_\nu, \\
\vec{j}_\nu &= -D_\nu\left[\nabla c_\nu + \frac{z_\nu e}{k_B\theta}c_\nu\nabla\phi \right],
\end{aligned}
\end{equation}
with $D_\nu$ the diffusion coefficient of ion $\nu$ in the bulk and interface
\begin{equation}
D_\nu(y) = \begin{cases}
    D_\nu^i, & \text{if $|y|<\ell$}.\\
    D_\nu^b, & \text{otherwise}.
  \end{cases}
\end{equation}
In addition, $z_\nu$ is the valence of the ions, and $\phi$ the electric potential field.

\begin{figure*}[htb]
\centering
\includegraphics[width=0.7\linewidth]{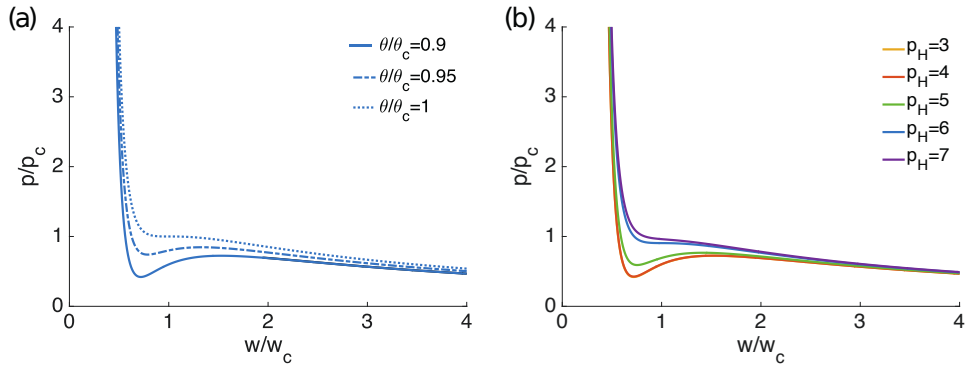}
 \caption{Projection onto the $p$--$w$ plane shows the sensitivity of the local state of the lipid to (a) temperature and (b) pH. The isotherms in (a) were calculated at an interfacial pH=4 (roughly the interfacial pH for typical lipids under physiological conditions and a bulk pH=7). The variation of pH in (b) was calculated at $\theta/\theta_c=0.9$. In both figures we have used pK=5, $c_H=c_{OH} = 10^{-4}$ mM, and $c_M=c_A =$ 10 and 100 mM on both sides of the interface, respectively.}
 \label{Fig-02}
 \end{figure*}

Electromagnetic fields obey Maxwell's equations. In the absence of a magnetic field, the electric field ($\vec{E}=-\nabla\phi$) obeys Gauss' and Ampere' laws
\begin{equation}\label{eq:Maxwell}
\begin{aligned}
\nabla \cdot \vec{E} &= \frac{\rho_{el}}{\varepsilon_0\varepsilon_r}, \\
\partial_t\vec{E} &= - \frac{\vec{J_{el}}}{\varepsilon_0\varepsilon_r},
\end{aligned}
\end{equation}
with the electric charge density and current being
\begin{equation}\label{eq:elecDensCurr}
\begin{aligned}
\rho_{el} &= eN_A\sum_\nu z_\nu c_\nu + \sum_\pm \sigma_\pm\delta(y\pm\ell),\\
\vec{J}_{el} &= eN_A\sum_\nu\vec{j}_\nu + \sum_\pm \partial_t\sigma_\pm\delta(y\pm\ell),
\end{aligned}
\end{equation}
and $\varepsilon_r$ is the relative permittivity of the bulk and interface
\begin{equation}\label{eq:dielectricConstant}
\varepsilon_r = \begin{cases}
    \varepsilon_r^i, & \text{if $|y|<\ell$}.\\
    \varepsilon_r^b, & \text{otherwise}.
  \end{cases}
\end{equation}
Conveniently, the conservation of mass (Eq.~(\ref{eq:NernstPlanck})) and Gauss' law (upper Eq.~(\ref{eq:Maxwell})) automatically satisfy Ampere' law (lower Eq.~(\ref{eq:Maxwell})). A detailed list of the variables and parameters of the model is provided in the Supplemental Materials (Tables S1--S2). 

Here we investigate only an idealized scenario of zero ionic permeability across the interface; i.e., $D_\nu^i = 0$ for all $\nu$. This simplification is reasoned by two arguments. Firstly, a lack of experimental data on the state dependent permeability of ions\cite{El-Mashak1985}. Secondly, by ignoring the effect of permeability on the transmembrane potential measurement, the large contribution of surface potential is highlighted. This was already pointed out by others\cite{ohki1980}, but is completely neglected in the classical electric treatment of APs\cite{hodgkin1949effect, Hodgkin1952}. In the limit of zero permeability across the interface, the ionic and electric variables are described by the classical Gouy-Chapman double layer potential, where positive ions accumulate near the negative surface charge and act to screen the electric potential at a distance $\sim10\bar{\lambda}_D$. The length scale $\bar{\lambda}_D$ is related to the Debye screening length (Eq.~(\ref{eq:lambdaDpm})), but unlike $\lambda_{D,\pm}$, it only contains constants of the model
\begin{equation}
\bar{\lambda}_D = \left(\frac{\varepsilon_0\varepsilon_r^b k_B\theta_c}{e^2 N_A \sum_{\nu,\pm} c_{\nu,\infty,\pm}}\right)^{1/2}.
\end{equation}

\begin{figure*}[htb]
\centering
\includegraphics[width=0.6\linewidth]{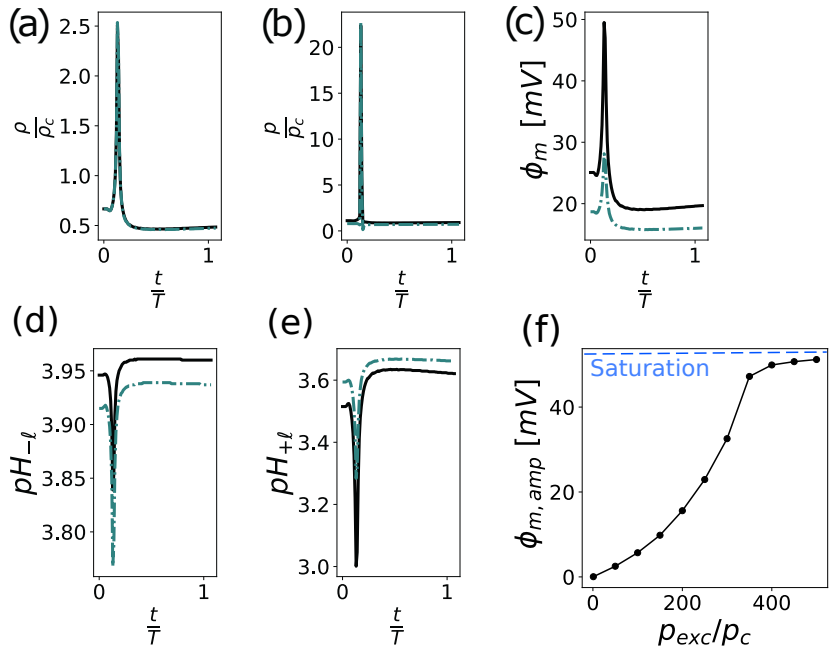}
 \caption{(a)--(e) Various aspects of an acoustic pulse that traverses the phase transition, as measured at a distance $x/L=1$ from the excitation point. The plots compare an asymmetric pK (solid black line) vs.~an asymmetry in the ionic concentration (dotted-dashed turquoise line). (a) Local density. (b) Pressure. (c) Transmembrane potential difference. (d) Local interfacial pH at $y=-\ell$ and at (e) $y=\ell$. Parameters: $\rho_0=0.67$, $pK_{-\ell}$=5, $pK_{\ell}$=3, $c_{M,\pm\infty}$=100 mM for solid line, and $pK_{\pm\ell}$=5, $c_{M,-\infty}$=100 mM, $c_{M,\infty}$=1 mM for dotted-dashed line. (f) Sigmoidal response of the voltage pulse amplitude to the amplitude of excitation. Parameters were $\rho_0=0.45$, $c_{M,\pm\infty}$=100 mM, and $pK_{-\ell}$=5, $pK_{\ell}$=7. Additional parameters that were used in all simulations: $pH_{\pm\infty}$ = 7. Excitation parameters were $(\tilde{x}_0, \tilde{t}_0, \tilde{p}_{exc}, \lambda)$=(0, 0.1, 250, 0.07). Numerical calculation was conducted with 1024 grid points, x-domain $[-5, 5]$, and $dt=10^{-3}$. Other parameters were similar to Fig.~2 of Ref.~\cite{Mussel2018}.}
 \label{Fig-03}
 \end{figure*}

\paragraph{System scales.} The critical point and the viscosity of the interface are used to define proper scales (time, length and velocity, respectively)
\begin{equation}\label{eq:scales}
T\equiv\frac{\zeta}{p_c} ,\quad L\equiv\zeta\sqrt{\frac{w_c}{p_c}} , \quad U\equiv\frac{L}{T}=\sqrt{w_c p_c}.
\end{equation}
These scales govern the longitudinal pulses that propagate parallel to the axis of the interface. Typical values, estimated from experiments with DPPC lipids, are T$\approx$30 ms, L$\approx$4 m and U$\approx$120 m/s \cite{Mussel2018}. In addition, the electric forces that act on the mobile ions define another set of time and length scales
\begin{equation}\label{eq:scales2}
T_D = \frac{\bar{\lambda}_D^2}{D_H^b}, \quad \bar{\lambda}_D
\end{equation}
where $D_H^b$ is the diffusion coefficient of protons in the bulk fluid. Typical values, estimated for physiological conditions ($D_H^b\sim10^{-9}~m^2/s$, $c_{H,\pm\infty}=c_{OH,\pm\infty}\sim 10^{-4}$ mM and $c_{M,\pm\infty}=c_{A,\pm\infty}\sim$10--100 mM), are $T_D\sim10^{-9}$ s and $\bar{\lambda}_D\sim10^{-9}$ m. The proper scales (Eqs.~(\ref{eq:scales})--(\ref{eq:scales2})) were used to define a dimensionless form of the model equations (Eqs.~(S13)--(S22) in the Supplemental Materials). For typical lipids in physiological conditions, the scales of acoustic pulses are much larger than the ionic response. Therefore, the concentration of ions and the electric potential may be considered in a momentary steady state (Eq.~S23). For the planar geometry considered here, a steady state solution of the ionic concentration and electric potential  was obtained analytically (Eqs.~(S26)--(S27)), and is sketched in Fig.~\ref{Fig-01}b. This is a typical Gouy-Chapmann solution, where positive ions congregate near the interface and exponentially reduce the electric potential. In the absence of ionic permeability the electric field inside the membrane is assumed negligible, and the transmembrane potential difference $\phi_m$ arises mainly from the fixed charges at the two leaflets and the surrounding electrolyte solution\cite{Ohki1971, Ohki1972}. 

The model equations were numerically solved with the Dedalus open-source code\cite{Burns2017}, which is based on a pseudo-spectral method. The model was solved using periodic boundary conditions at the x-axis, and with homogeneous initial conditions ($\rho_0, p_0, \theta_0, c_{\nu, \pm\infty}$). Pulses were excited by locally increasing the stress for a brief period of time; i.e., the following term was added into the right-hand-side of the middle Eq.~\ref{eq:consLaws}
\begin{equation}\label{eq:excitationCurrent}
\bar{w}\partial_h \left(p_{exc} \Theta(t_0-t ) e^{-\frac{(h-h_0)^2}{2\lambda^2}} \right).
\end{equation}
Here, $p_{exc}$ is the amplitude of excitation, $\Theta$ is the Heaviside function, $t_0$ is the duration of the excitation, and $h_0$ and $\lambda$ are the spatial center and width of the excitation. 

\section{Results}
\begin{figure*}[t]
\centering
\includegraphics[width=0.9\linewidth]{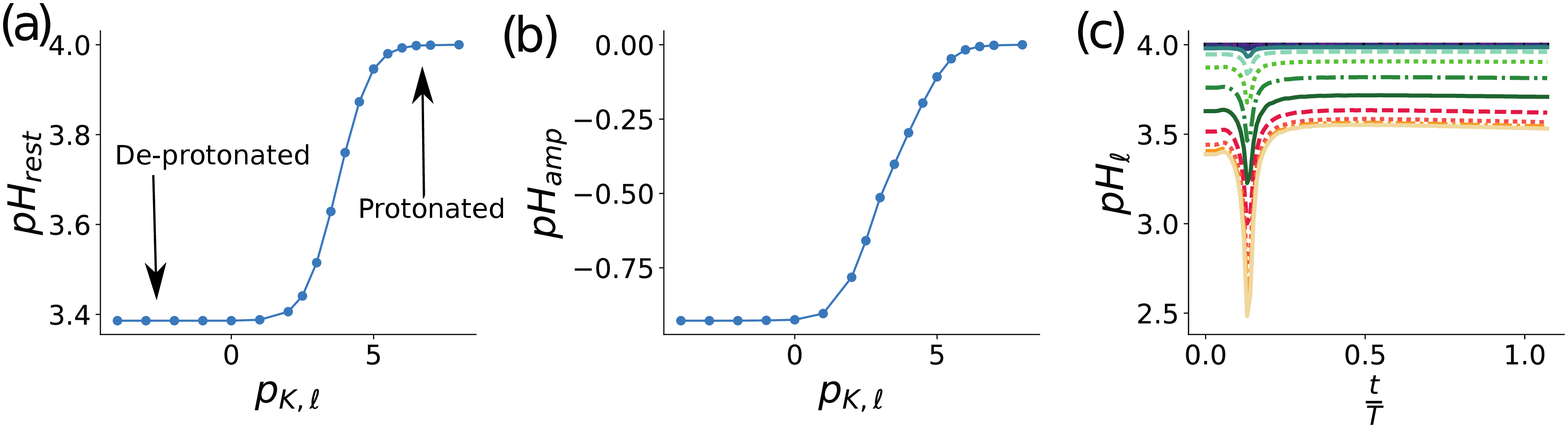}
 \caption{The effect of pK on the pH pulse . (a) resting pH. (b) Amplitude of pH pulse. (c) pH pulse as a function of time as measured at distance $x/L=1$ from the excitation point. Multiples pulses are plotted in rainbow colors, ranging from $pK_\ell$=7 (dark-purple) to $pK_\ell$=0 (orange). Parameters are similar to Fig.~\ref{Fig-03}a--e.}
 \label{Fig-06}
 \end{figure*}

\paragraph{vdW--Tr{\"a}uble state diagram.} First, let us explore the constitutive equation of the interface, as described by the combined van der Waals and Tr{\"a}uble models. The model proposed by Tr{\"a}uble captures the effect of ionic concentration on the state of the interface. For simplicity we considered only the response to pH, and ignored interaction with other mobile particles. Hence, we set $K_\nu=0$ for $\nu\neq H^+$. In this case, the critical point is no longer a zero dimensional point, but rather a pH dependent ``critical line''. The model qualitatively captures the response of the lipid state to temperature and density variation\cite{Albrecht1978}. This is demonstrated in Fig.~\ref{Fig-02}a, where a projection into the $p$--$w$ plane is given for three values of temperature, at a constant interfacial pH=4. As is shown below (Fig.~\ref{Fig-03}d,e and Fig.~\ref{Fig-06}), this is roughly the interfacial pH for typical lipids under physiological conditions and a bulk pH of 7. Figure \ref{Fig-02}b demonstrates the effect of pH on the local state of the lipid interface. The pressure is particularly sensitive to pH variations when the pH is sufficiently close to the $pK$ (roughly less than two scale units), and has almost no effect for pH values far from it. This behavior results from the partial screening of the lipid charged sites, and qualitatively captures previous experimental results\cite{trauble1977membrane, Fichtl2016}. We therefore conclude that the combined van der Waals and Tr{\"a}uble's models qualify as a first order approximation of the state relation between the pressure, density, temperature and pH at the interface.

\paragraph{Scales and asymmetry of the electric properties.} Heimburg and Jackson have previously suggested that density variations within the lipid interface generate changes in the surface potential\cite{Heimburg2005}. Although we in principle agree with their reasoning, their predictions are based on a small amplitude approximation of both the density and the electric potential. Furthermore, the scale of the electric potential was not derived from the theory, but rather was estimated from past experiments ($\approx$100 mV). Finally, their electrical potential estimation does not consider the symmetry/asymmetry of the lipid bilayer. The importance of the latter was previously discussed by Ohki\cite{Ohki1971, Ohki1972}, and is further demonstrated here. 

\begin{figure*}[htb]
\centering
\includegraphics[width=0.8\linewidth]{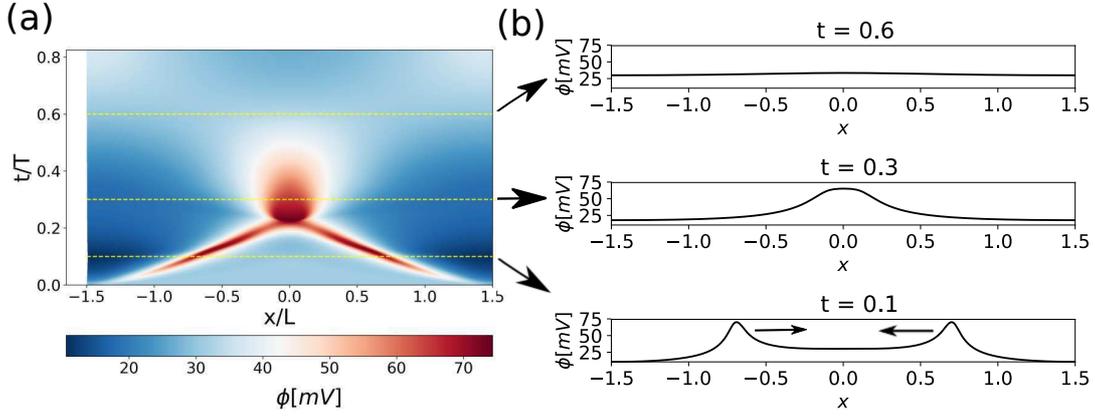}
 \caption{Annihilation of pulses upon collision, as observed by the transmembrane potential difference. (a) Complete x-t solution of the voltage field. (b) Snapshot of the transmembrane potential difference field before (t/T=0.1), during (t/T=0.3), and after the collision (t/T=0.6). Parameters: $\rho_0=0.67$, $pK_{-\ell}$=5, $pK_{\ell}$=2, $c_{M,\pm\infty}$=100 mM, and x-domain=$[-1.5, 1.5]$. Other parameters were similar to Fig.~\ref{Fig-03}a--e.}
 \label{Fig-04}
 \end{figure*}
 
The proper scale of the electric potential, as arises from the model, is $\phi_0 = k_B\theta_c/e\approx$-25 mV (Eq.~(S14)). Furthermore, the difference in density between the lipid-expanded and the liquid-condensed phases results in a difference in surface potential of $\approx$5--50 mV (Eq.~(S27)). This is only slightly smaller than measurements in lipid monolayers ($\approx$100 mV)\cite{Steppich2010}. However, when estimating a transmembrane potential, as detected by distant electrodes, it is important to consider the symmetry/asymmetry of the membrane. The reason is that irrespective of the state of the interface the electric potential is completely screened by the mobile ions at a distance $\sim10\bar{\lambda}_D\approx$10 nm (Fig.~\ref{Fig-01}b). This implies that for a symmetric system with respect to the axis of the interface (x-axis in Fig.~\ref{Fig-01}a), changes at the interface cannot be detected by electrodes in the bulk. Nonetheless, the introduction of an asymmetry; for example, a difference in the ionic concentration on both sides of the interface, results in a nonzero electric potential difference, $\phi_m$ (Fig.~\ref{Fig-01}b and Eq.~(S29)). Moreover, the transmembrane potential difference is sensitive to state changes at the interface, and provides a useful and simple access to detecting traveling acoustic pulses from afar. 

\begin{figure*}[b]
\centering
\includegraphics[width=1.0\linewidth]{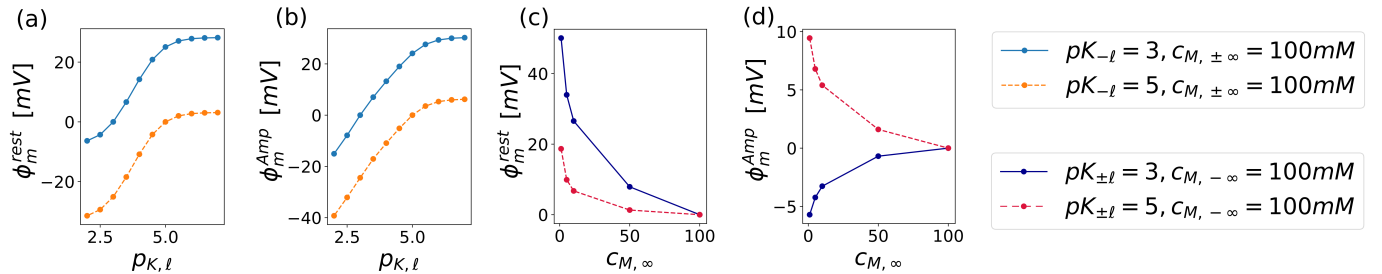}
 \caption{The effect of pK and ionic concentration asymmetry on the resting potential and electric pulse amplitude as measured at distance $x/L=1$ from the excitation point. (a) Variation of the resting potential with pK at one leaflet while keeping the pK at the other leaflet constant at value of 3 (blue solid line) or 5 (orange dashed line). (b) Dependence of the pulse amplitude on the pK. (c) Variation of the resting potential with the ionic concentration difference. The concentration was $c_{M,-\infty}=c_{A,-\infty}$=100 mM at one side of the interface, and varied at the other. Dark-blue solid line represents pK=3 at both sides of the interface and red dashed line represents pK=5. (d) effect of ionic concentration asymmetry on the amplitude of the electric pulse. Parameters are similar to Fig.~\ref{Fig-03}a--e.}
 \label{Fig-05}
 \end{figure*}

\paragraph{The mechanical aspect of acoustic pulses.} We now turn to explore solutions of acoustic pulses by numerically solving the coupled model equations (Eqs.~(\ref{eq:consLaws}), (\ref{eq:constitutiveEqs}) (\ref{eq:NernstPlanck}), and (\ref{eq:Maxwell})). Figure \ref{Fig-03}a--e depicts various aspects of an acoustic pulse that traverses the phase transition for two scenarios, one with an asymmetry in the pK of the leaflets (solid black line) and another with an asymmetry in the ionic concentration on both sides of the interface (dotted-dashed turquoise line). Figure \ref{Fig-03}a,b shows the density and pressure aspects, respectively, as measured at a distance $x/L$=1 from the excitation point. These aspects acted similarly in both scenarios, and were described in details in our previous work\cite{Mussel2018}. In brief, a sharp rise in density appeared as the local region of the interface traversed the phase transition, and was followed by a sharp rise in pressure when the state reached the liquid-condensed phase. Subsequently, the local state relaxed in a reverse order, first in pressure and than in density. These mechanical aspects were accompanied by an adiabatic change of temperature (data not shown), and in addition by variations in the transmembrane potential difference (Fig.~\ref{Fig-03}c) and surface pH at the two leaflets (Figs.~\ref{Fig-03}d,e). These are described in more details below.

\paragraph{The electrical aspect of acoustic pulses} The shape of the transmembrane potential pulse (Fig.~\ref{Fig-03}c) resembles the typical shape of an AP, which is characterized by a recognizable depolarization, repolarization and hyperpolarization regions\cite{Hodgkin1952}. Moreover, the scales of the electric pulse, for typical lipid values, are of similar orders of magnitude as APs; the resting potential and pulse amplitude are $\sim$$\phi_0$$\sim$10 mV and the pulse duration is $\sim$0.1T$\sim$1 ms. Figure \ref{Fig-03}f demonstrates that by increasing the amplitude of excitation, the amplitude of the voltage pulse shows a sigmoidal response, which resembles the density response to excitation\cite{Mussel2018, Shrivastava2015} as well as the ``all-or-none'' response of APs\cite{aidley1998physiology}. This non-linear behavior follows from the phase transition and volume exclusion regions. At small amplitudes of excitation the local state of the lipid remains in the lipid-expanded phase, and the response to excitation is almost linear. For larger amplitudes of excitation, the local state penetrates the phase transition region, and the density (as well as the accompanying electric signal) increases with almost no penalty of pressure. For even larger amplitudes of excitation, the local state changes into the liquid condensed state, and the density saturates due to the excluded volume of the lipid molecules. Figure \ref{Fig-04} further demonstrates that these non-linear pulses annihilate upon collision for the parameters that describe typical lipids\cite{Mussel2018}. This property, too, is similar to APs\cite{Tasaki1949}. Interestingly, but beyond the scope of the manuscript, the model displays a rich behavior of response to collision in other regions of the parameter space.

The characteristics of the transmembrane potential pulse were investigated for two types of asymmetries; interfacial pK and ionic concentration. Upon fixing the pK at one leaflet and increasing the pK of the other leaflet, the resting potential monotonically increases and ranges between $\approx$-30 and 30 mV (Fig.~\ref{Fig-05}a). It only becomes zero when the reflection symmetry is restored. Importantly, the calculation was conducted for a system in the absence of ionic concentration gradient and permeability. This example demonstrates that the electric charge at the interface and its association to mobile particles cannot be ignored, as was previously argued\cite{ohki1980} and in contrast to the classical treatment of the resting potential\cite{hodgkin1949effect, Hodgkin1952}. 

The amplitude of the voltage aspect varies too upon changing the pK at a leaflet, as shown in Fig.~\ref{Fig-05}b for a scenario in the absence of ionic gradients. A crucial prediction, therefore, is that transmembrane potential pulses may be detected in the absence of ionic gradients, as long as the state of the asymmetric interface remains sufficiently close to the phase transition. This prediction is substantially different from the classical electrical theory of APs\cite{Hodgkin1952}. In a support to the prediction, it was previously demonstrated that neurons can develop a repetitive voltage response, that resembles APs in shape and scales, in the complete absence of ionic gradients\cite{Terakawa1981}. 

The electric properties of the interface are further affected by an asymmetry of the bulk ionic content. Figure \ref{Fig-05}c shows the effect of the monovalent salt concentration on the resting potential. A difference in ionic concentration at physiological scales ($\sim$1--100 mM) gives rise to a significant resting potential ($\sim$10 mV), that monotonically increases upon increasing the asymmetry. The amplitude of the electric aspect of acoustic pulses, as generated solely by the ionic asymmetry, is of slightly smaller scale, $\sim$1 mV (Fig.~\ref{Fig-05}d). It is worthwhile to mention that the asymmetry between the two leaflets has to be considered the rule rather than an exception. Living cells, for example, are asymmetric from the get go, and irrespective of active features (e.g, pumps). 
 
\paragraph{The pH aspect of acoustic pulses.} The local pH at the interface is strongly influenced by the surface charge and pK of the leaflet. The sigmoidal effect of the latter is shown in Fig.~\ref{Fig-06}a. It demonstrates a transition from completely protonated leaflet, when the pK is larger by more than 2 units than the resting pH, to completely de-protonation, when the pK is smaller by more than 2 units than the resting pH. Acoustic pulses generate a local variation in the interfacial pH (Figs.~\ref{Fig-06}b,c), with an amplitude that strongly depends on the pK of the leaflet. The pH pulse amplitude is maximum when the leaflet is completely de-protonated, and decreases by almost one unit for typical lipid values. In contrast, when the leaflet is fully protonated, the amplitude of the pH pulse becomes negligible. This happens because the negatively charged sites at the leaflet are saturated with protons already in the lipid-expanded phase, and the longitudinal density changes hardly affect the local concentration of protons. These results are in accord with experimental observations in lipid monolayers\cite{Fichtl2016, Fichtl2018}.

 \section{Discussion}
APs share many similar properties with acoustic pulses that traverse the melting transition in lipid interfaces\cite{Shrivastava2014, Shrivastava2015, Shrivastava2017, Mussel2018}. These include time and velocity scales, the appearance in multiple observables, an adiabatic temperature signal, a sigmoidal response to excitation, and annihilation upon collision. In addition, it was previously demonstrated that the membrane potential and the pH are locally affected by acoustic pulses in lipids\cite{Steppich2010, Fichtl2016}. In this work, we have continued this line of research, and demonstrated that electrical and pH aspects of acoustic pulses emerge from an idealized physical description of the lipid interface. The most significant result of the model is that the transmembrane potential pulse is similar to APs both in shape (distinct depolarization, repolarization and hyperpolarization regions) and scales (millivolts and milliseconds). 

\paragraph{Cellular excitability.} Many types of living cells, from all major biological taxa, are excitable given the suitable conditions (see references within Ref.~\cite{Mussel2018}). These findings are naturally explained by the acoustic picture, where excitability is related to the state diagram of the interface. Specifically, when the state of the interface is sufficiently far from a phase transition region, it is more difficult to generate and detect these types of pulses, as was previously argued\cite{Kaufmann1989, Heimburg2005, Shrivastava2014}. In addition, the observation that APs can be excited by multiples means (mechanical stress, temperature, pH, electric currents and electromagnetic waves) is also natural from the thermodynamic picture, and does not require specific ``molecular receptors''\cite{Kaufmann1989, Heimburg2005, Mussel2018}. 

The work presented here offers additional possibilities that may explain the rich phenomenology of excitable cells. For example, the pH aspect of acoustic pulses can be sensed by other interfaces in close proximity, $\lesssim$ 10 nm. This distance closely resembles the synaptic cleft, where neighboring neurons interact. Particularly, as acoustic pulses follow from fundamental physical principles, it cannot be neglected that a variation of the acid concentration at the synapse excites an acoustic pulse in the post-synaptic cell, similar (or instead) of receptor activation\cite{fillafer2016excitation}. On the other hand, we have also demonstrated that when the interface is completely protonated, an acoustic pulse will generate a negligible pH variation (Fig.~\ref{Fig-06}). In this scenario, pulses may be detected electrically in the pre-synaptic cell, but may not activate any response from the post-synaptic cell. An increase of the interfacial pK may be another possible mechanism for the action of toxins and pharmacological agents (in addition to shifting the state of the interface away from the phase transition region\cite{Kaufmann1989, Heimburg2005, Mussel2018}).

The conditions of excitability of acoustic pulses are more convoluted than a threshold picture, and depend on the local state of the interface. This has a direct implication on information processing, typically associated with pulses in excitable cells. Such cells are sometimes treated as computational elements, that form the basis of behavioral organisms. According to the classical computational scheme, excitable cells fire an AP only when the membrane potential reaches a certain threshold, and therefore convey binary data to successor cells. In contrast, acoustic pulses transmit additional information, for example, by the pressure aspect. In contrast to the density, pH and voltage aspects, the pressure does not saturate, but rather increases monotonically with the excitation amplitude\cite{Mussel2018}. This analogue mechanical variation may propagate to neighboring cells via connecting protein complexes and elicit a graded response.  Thus, the possibility that APs are acoustic pulses should dramatically modify this ``digital-like'' picture, both in the content of information propagated as well as the state-dependent cellular response

\paragraph{Model extensions.} The model presented in this work illustrates how a synthesis of multiple physical aspects (mechanical, thermal, electrical and chemical) results in a rich description that semi-quantitatively captures biological phenomenology. Still, some aspects of the model are rather idealized. Here we list several aspects that should be addressed in a future work to produce a more accurate description of acoustic pulses in soft interfaces. (1) The hydrodynamics of the bulk was completely ignored in this work, and can be restored by considering mass, momentum and energy conservation laws at the bulk. Specifically, the bulk influences the pulse properties via the viscous coupling to the interface (e.g., it reduces the propagation velocity)\cite{Griesbauer2012, Kappler2017}. In addition, incorporation of the bulk dynamics will allow a more accurate estimation of the temperature aspect, as detected by distant sensors\cite{Tasaki1999}. (2) The magnetic field was also neglected in this work and can be restored by considering the complete Maxwell's equations.  Therefore, magnetic changes are unavoidable during acoustic pulses, and were indeed detected during APs\cite{Wikswo1980}.  (3) In addition to temperature and pH, the state of the lipid interface is sensitive to other thermodynamic quantities, including additional ionic components\cite{Trauble1974} as well as external electric fields\cite{antonov1990electric}. These effects can be easily incorporated into the model via the constitutive equation of the interface (which is directly related to its free energy). (4) The model contains many constants, which in reality are thermodynamic; i.e., state dependent. These include the heat capacity, thermal conductivity, interfacial viscosity, association constant with the protons, dielectric constant as well as the permeability to mobile particles. (5) The two leaflets of the interface were treated in this work as having separate charge density, while keeping other variables mutual (e.g., local pressure and density). It is, however, more likely that the two leaflets should be treated as separate, but coupled systems. This will allow, for example, for one leaflet to be in the liquid-expanded state while the other is in the liquid-condensed state. It was previously suggested that such a scenario results in oscillations of pH and transmembrane potential, the latter resembles a train of action potentials\cite{Yagisawa1993}.

\section{Conclusions}
We have theoretically demonstrated that electrical and chemical changes are an inseparable part of acoustic pulses in lipid interfaces. The description of the material is based on classical physical principles that must exist in lipids (e.g., conservation laws), and without adjustable fit parameters. Particularly, our theory uses an extension of the van der Waals model that includes electrical properties (a vdW--Tr{\"a}uble constitutive equation). The local interfacial pressure, therefore, depends on the electric potential and the concentration of nearby mobile ions, in addition to the density and temperature, $p(w, \theta, \phi, c_\nu)$. Thus, the constitutive equation serves as the electro-mechano-thermo-chemical coupling of the material
 
 The theory further reveals that upon breaking the reflection symmetry with respect to the interface, a nonzero transmembrane potential difference arises in the bulk, and may be captured by distant electrodes. Importantly, due to the thermodynamic coupling of the interface ($p(w, \theta, \phi, c_\nu)$), any asymmetry will lead to the emergence of the electric potential difference. Since the potential difference depends on the local state of the interface, it provides direct information on the propagation of acoustic pulses at the interface. 
 
The model complements a previous work, where the unusual properties of acoustic pulses that traverse the phase transition were obtained\cite{Mussel2018}. Taken together, we have demonstrated that these pulses show striking similarities to action potentials, including the qualitative shape of the pulse, the time, velocity and voltage scales, the saturation of pulse amplitude at strong excitations, as well as annihilation upon collision. Our theory provides a physical basis to describe acoustics in living and non-living soft systems, and supports the hypothesis of Kaufmann, Heimburg, and Jackson, that action potentials are of an acoustic nature\cite{Kaufmann1989, Heimburg2005}.

\footnotesize
\section{Acknowledgements}
The authors thank Christian Fillafer for fruitful discussions. MM further thanks Omri Har-Shemesh for valuable comments, and Daniel Lecoanet and Jeffrey S.~Oishi for support with the Dedalus open-source code. MM also acknowledges funds from SHENC-research unit FOR 1543.

\section{Author contributions}
MM and MFS devised the project and the main conceptual ideas. MM performed the computations. MM and MFS wrote the manuscript.

\section{Additional information}
{\bf Competing interests:} The authors declare that they have no competing interests.

\end{multicols}


\rule{\textwidth}{0.5pt}

\begin{multicols}{2}
{
\footnotesize

}
\end{multicols}

\end{document}


\maketitle

\begin{multicols}{2}

\beginsupplement

\begin{appendices}

The supplemental materials are intended to provide further physical and mathematical details, including Tra{\"u}ble's model on the electrostatic contribution to the membrane interfacial pressure (section \ref{sec:Trauble}), and further details on the model equations (section \ref{sec:equations}).


\section{Electrostatic effects on membrane structure}\label{sec:Trauble}
Adsorption or desorption of proton by the polar heads of acidic lipid molecules can trigger the melting phase transition in lipid interfaces [14]. In a seminal work, Hermann Tra{\"u}ble quantified the electric effect of protons on lipids by calculating the free energy contribution of the  electrostatic interactions [12]. Here we briefly repeat Tra{\"u}ble's assumptions and mathematical expressions. 

The electrostatic contribution to the specific free energy of a charged surface is calculated from the charging process of the surface along the pathway of equilibrium states
\begin{equation}
f_{el} = w\int_0^{\sigma} \phi_0(\sigma) d\sigma.
\end{equation}
Here $w$ is the specific volume of the interface, $\phi_0$ is the electric potential at the plane of the fixed interfacial charges, and $\sigma$ is the surface charge density. For a planar geometry one can obtain analytic expression using the Poisson equation
\begin{equation}
\partial_x^2 \phi = -\frac{eN_A}{\varepsilon_0\varepsilon_r^b}\sum_\nu z_\nu c_{\nu,\infty}e^{\frac{-z_\nu e\phi}{k_B\theta}},
\end{equation}
with $x$ the dimension perpendicular to the surface. Using the identity
\begin{equation}
\partial_x^2\phi = \frac{1}{2}\partial_\phi\left(\partial_x\phi\right)^2
\end{equation}
and integrating over $\phi$, we obtain
\begin{equation}
\partial_x\phi = \sqrt{2\frac{k_B\theta}{\varepsilon_0\varepsilon_r^b}N_A\sum_\nu c_{\nu,\infty}\left(e^{\frac{-z_\nu e\phi}{k_B\theta}} - 1\right)}.
\end{equation}
The surface charge density is related to the slope of the electric potential at the plane of the fixed charges, $\partial_x\phi |_{surface} = \sigma/\varepsilon_0\varepsilon_r$. Hence, $\phi_0$ is related to $\sigma$ according to  
\begin{equation}
\sigma= \sqrt{2\varepsilon_0\varepsilon_r^b k_B\theta N_A\sum_\nu c_{\nu,\infty}\left(e^{\frac{-z_\nu e\phi_0}{k_B\theta}} - 1\right)}.
\end{equation}

In the work presented here we have considered a simplified description that includes only four monovalent ions (proton $H^+$, hydroxide $OH^-$, one type of alkaline ion $M^+$, and one type of halogen ion $A^-$). For this 1:1:1:1 electrolyte, and using the hyperbolic identity $\cosh(2x) - 1 = 2\sinh^2(x)$, the expression is simplified into
\begin{equation}
\sigma= \sigma_D\sinh\left(\frac{e\phi_0}{2k_B\theta}\right),
\end{equation}
with
\begin{equation}
\sigma_D = \frac{4\varepsilon_0\varepsilon_r^b k_B\theta}{e\lambda_D}, 
\end{equation}
and $\lambda_D$ is the Debye length
\begin{equation}
\lambda_D = \left(\frac{\varepsilon_0\varepsilon_r^b k_B\theta}{e^2N_A\sum_\nu c_{\nu,\infty}}\right)^{1/2}.
\end{equation}
In this idealized system, the specific free energy can be solved analytically
\begin{equation}
f_{el} = w\left(\sigma\phi_0 - \frac{2k_B\theta\sigma_D}{e}\left[\sqrt{1+\left(\frac{\sigma}{\sigma_D}\right)^2} - 1\right] \right),
\end{equation}

To calculate the electrostatic contribution to the interfacial pressure, $p_{el} = -\left(\partial f_{el}/\partial w\right)_\theta$, we recall that the interfacial charge density depends on the specific volume as well as the binding to ions (here we repeat Eq.~(6))
\begin{equation}
\sigma = -\frac{e}{mw}\frac{1}{1+\sum_\nu K_\nu c_\nu}.
\end{equation}
This relation implies that $\partial_w\sigma = -\sigma/w$. Following some algebraic manipulations, the expression for the interfacial pressure is
\begin{equation}
p_{el} = \frac{2k_B\theta\sigma_D}{e}\left[\sqrt{1+\left(\frac{\sigma}{\sigma_D}\right)^2} - 1\right].
\end{equation}

\section{Model equations}\label{sec:equations}
\paragraph{Model variables and parameters.} The system considered in this letter is described by ten dynamic variables of space and time, that are coupled to one another by three conservation laws of the interface (Eq.~(1)), two constitutive equations (Eq.~(4)), four conservation laws of the mobile ions (Eq.~(9)) and Gauss' law of electric fields (upper Eq.~(10)). The ten variables are listed in Table \ref{tab:variables}. In addition, the model depends on measurable parameters that characterize the interface, the mobile ions and the electric field (Table \ref{tab:parameters}). The van der Waals parameters ($m, a, b$) are related to the critical point according to
\begin{equation}
w_c=\frac{3b}{m}, \qquad p_c=\frac{a}{27b^2}, \qquad \theta_c=\frac{8a}{27b k_B}.
\end{equation}
The scaling of the spatial coordinate $h$ (in the Lagrange frame) was defined according to the critical density ($\bar{w}=w_c$). Finally, $k_B\cong 1.38\cdot 10^{-23}~J/K$ is the Boltzmann constant,  $\varepsilon_0=8.85\cdot 10^{-12}~C^2/J\cdot m$ the vacuum permittivity, $e=1.6\cdot10^{-19}~C$ is the elementary electric charge, $z_\nu=\pm1$ is the ionic valence, and $N_A=6.02\cdot 10^{23}$ is Avogadro number.

{
\renewcommand{\arraystretch}{1.5}
\begin{table*}
\centering
\begin{tabular}{ | c | c | p{3.5cm} | }
\hline
Variable & Name & Units for a 3d material (2d material) in MKS system \\ \hline
$\rho=w^-1$ & density & $\frac{m^3}{kg}~\left(\frac{m^2}{kg}\right)$ \\ \hline
$v$ & velocity & $\frac{m}{s}$ \\ \hline
$E$ & specific total energy & $\frac{m^2}{s^2}$ \\ \hline
$p$ & pressure & $\frac{N}{m^2}~\left(\frac{N}{m}\right)$ \\ \hline
$\theta$ & temperature & $K$ \\ \hline
$\phi$ & electric potential & $V$ \\ \hline
$c_\nu$ & ionic concentration & $\frac{mole}{m^3}$ \\ \hline
 \end{tabular}
 \caption{Model variables}
 \label{tab:variables}
 \end{table*}
}

{
\renewcommand{\arraystretch}{1.5}
\begin{table*}
\centering
\begin{tabular}{ | c | c | p{3.5cm} | }
\hline
Parameter & Name & Units for a 3d material (2d material) in MKS system \\ \hline
$m$ & mass of a fluid particle & $kg$ \\ \hline
$a$ & average attraction between fluid particles & $Nm^4 \left(Nm^3\right)$ \\ \hline
$b$ & volume exclusion by the particles & $m^3 \left(m^2\right)$ \\ \hline
$\zeta$ & dilatational viscosity & $\frac{kg}{m\cdot s}~\left(\frac{kg}{s}\right)$ \\ \hline
$c_v$ & specific heat capacity at constant volume & $\frac{J}{kg\cdot K}$ \\ \hline
$k$ & thermal conductivity & $\frac{J}{m\cdot s\cdot K}~\left(\frac{J}{s\cdot K}\right)$ \\ \hline
$C$ & capillarity coefficient & $\frac{kg^2}{m^2\cdot s^2}~\left(\frac{kg^2}{s^2}\right)$ \\ \hline
$K_{\nu,\pm}$ & binding constant of ion $c_\nu$ to ionized lipids & $\frac{m^3}{mole}$ \\ \hline
$D_{\nu}(y)$ & diffusion coefficient of ion $c_\nu$ in the bulk and interface & $\frac{m^2}{s}$ \\ \hline
$\varepsilon_r(y)$ & relative permittivity of the bulk and interface & - \\ \hline
$c_{\nu,\pm,\infty}$ & ionic concentration sufficiently far from the interface & $\frac{mole}{m^3}$ \\ \hline
 \end{tabular}
 \caption{Model constant parameters}
 \label{tab:parameters}
 \end{table*}
}

\paragraph{Dimensionless model equations.} The proper scales (Eqs.~(13)--(14)) were used to introduce the following reduced coordinates
\begin{equation}
\begin{aligned}
X\equiv \frac{h}{L},& \quad \tilde{x}\equiv \frac{x}{L}, \quad \tilde{y}\equiv\frac{y}{\bar{\lambda}_D}, \quad \tilde{t}\equiv \frac{t}{T} \\
& \tilde{\delta}(\tilde{y}) \equiv\bar{\lambda}_D\delta(y).
\end{aligned}
\end{equation}
and the spatial dimension in the Lagrange frame (Eq.~(2)) was normalized by the choice $\bar{w}=w_c$. The dimensionless model variables are
\begin{equation}\label{eq:dimlessVars}
\begin{aligned}
\tilde{w}\equiv\frac{w}{w_c}, \quad & \tilde{v}\equiv \frac{v}{U}, \quad \tilde{E}\equiv\frac{E}{U^2},   \\
\tilde{\theta}\equiv\frac{\theta}{\theta_c}, \quad &\tilde{p}\equiv\frac{p}{p_c}, \quad \tilde{\tau}\equiv\frac{\tau}{p_c}, \\
\tilde{\sigma}_\pm \equiv \frac{mw_c}{e}\sigma_\pm, \quad &\tilde{c}_{\nu,\pm} \equiv \bar{\lambda}_D^3 N_A c_{\nu,\pm}, \\
\tilde{\phi}\equiv\frac{k_B\theta_c}{e}\phi, \quad & \frac{\lambda_{D,\pm}}{\bar{\lambda}_D}  = \left(\tilde{\theta}\frac{\sum_{\nu,\pm} c_{\nu,\pm\infty}}{\sum_{\nu} c_{\nu,\pm\infty}}\right)^{1/2},
\end{aligned}
\end{equation}
and the model parameters are
\begin{equation}\label{eq:dimlessParam}
\begin{aligned}
\tilde{c}_v=\frac{c_v \theta_c}{p_c w_c}, \quad \tilde{k}&=\frac{k\theta_c}{p_c w_c \zeta},\quad \tilde{C}=\frac{C}{\zeta^2} \\
\tilde{K}_{\nu,\pm}=\frac{K_{\nu,\pm}}{N_A\bar{\lambda}_D^3}, &\quad \tilde{D}_\nu=\frac{D_\nu}{D_W}.
\end{aligned}
\end{equation}
Two additional dimensionless parameters are
\begin{equation}
\alpha\equiv\frac{e^2}{k_B\theta_c\varepsilon_0\bar{\lambda}_D}, \quad \chi\equiv\frac{\bar{\lambda}_D^2}{3b}.
\end{equation}

The dimensionless model equations are
\begin{equation}\label{eq:dimlessEqs}
\begin{aligned}
\partial_{\tilde{t}}  \tilde{w}&=\partial_X \tilde{v}, \\
\partial_{\tilde{t}}  \tilde{v}&=\partial_X \left(\tilde{\tau}_1+\tilde{\tau}_2 \right),  \\
\partial_{\tilde{t}}  \tilde{E}&=\partial_X \left(\tilde{\tau}_1 \tilde{v} \right)+\tilde{k}\partial_X^2 \tilde{\theta},
\end{aligned}
\end{equation}
with
\begin{equation}\label{eq:stress}
\begin{aligned}
\tilde{\tau}_1 &=-\tilde{p} + \partial_X \tilde{v}, \\
\tilde{\tau}_2 &= -\tilde{C}\partial_X^2\tilde{w},
\end{aligned}
\end{equation}
and
\begin{equation}\label{eq:dimlessvdW}
\begin{aligned}
\tilde{p}&=\frac{8\tilde{\theta}}{3\tilde{w}-1}-\frac{3}{\tilde{w}^2} + \tilde{p}_{tr,+} + \tilde{p}_{tr,-}, \\
\tilde{E}&=\frac{\tilde{v}^2}{2} + \tilde{c}_v \tilde{\theta}-\frac{3}{\tilde{w}} + \frac{\tilde{C}}{2}(\partial_X\tilde{w})^2 + \tilde{w}\left(\tilde{p}_{tr,+} + \tilde{p}_{tr,-}\right),
\end{aligned}
\end{equation}
with
\begin{equation}\label{eq:dimlessPtr}
\tilde{p}_{tr,\pm} = \frac{16}{3} \frac{4\chi}{\alpha}\frac{\bar{\lambda}_D\varepsilon_r^b\tilde{\theta}^2}{\lambda_{D,\pm}} \left[\sqrt{1+\left(  \frac{\alpha}{4\chi}\frac{\lambda_{D,\pm}}{\bar{\lambda}_D}  \frac{\tilde{\sigma}_\pm}{\varepsilon_r^b\tilde{\theta}} \right)^2}-1 \right].
\end{equation}

The ionic mass conservation is
\begin{equation}\label{eq:DimLessMassCons}
\begin{aligned}
\frac{T_D}{T}\partial_{\tilde{t}}\tilde{c}_\nu = \left(\frac{\bar{\lambda}_D}{L}\right)^2\tilde{D}_\nu\partial_{\tilde{x}}\left(\partial_{\tilde{x}}\tilde{c}_\nu + \frac{z_\nu}{\tilde{\theta}}\tilde{c}_\nu\partial_{\tilde{x}}\tilde{\phi}\right) \\
+ \partial_{\tilde{y}}\left[\tilde{D}_\nu\left(\partial_{\tilde{y}}\tilde{c}_\nu + \frac{z_\nu}{\tilde{\theta}}\tilde{c}_\nu\partial_{\tilde{y}}\tilde{\phi}\right)\right]
\end{aligned}
\end{equation}
and Gauss' law is
\begin{equation}\label{eq:dimLessGaussLaw}
 \left(\frac{\bar{\lambda}_D}{L}\right)^2\partial_{\tilde{x}}^2\tilde{\phi} + \partial_{\tilde{y}}^2\tilde{\phi} = -\frac{1}{\alpha\varepsilon_r}\left[\sum_\nu z_\nu \tilde{c}_\nu + \chi\sum_\pm\tilde{\sigma}_\pm\tilde{\delta}(\tilde{y}\pm\ell/\lambda_D) \right].
\end{equation}
The ionic response in the bulk fluid is typically much faster than the scale of acoustic pulses in typical lipids ($T_D/T\sim$$10^{-4}$--$10^{-3}$ and $\bar{\lambda}_D/L\sim$$10^{-9}$--$10^{-6}$). Therefore, at the bulk fluid Eqs.~(\ref{eq:DimLessMassCons})--(\ref{eq:dimLessGaussLaw}) can be simplified into
\begin{equation}\label{eq:simpGauss}
\begin{aligned}
0 &= \partial_{\tilde{y}}\left[\tilde{D}_\nu\left(\partial_{\tilde{y}}\tilde{c}_\nu + \frac{z_\nu}{\tilde{\theta}}\tilde{c}_\nu\partial_{\tilde{y}}\tilde{\phi}\right)\right], \\
\partial_{\tilde{y}}^2\tilde{\phi} &= -\frac{1}{\alpha\varepsilon_r}\left[\sum_\nu z_\nu \tilde{c}_\nu + \chi\sum_\pm\tilde{\sigma}_\pm\tilde{\delta}(\tilde{y}\pm\ell/\lambda_D) \right].
\end{aligned}
\end{equation}
It is important to note that within the interface the time derivative in Eq.~(\ref{eq:DimLessMassCons}) might not be negligible, because of a low value of the diffusion coefficient.

\paragraph{Analytic solution of the electric potential and ionic concentration.} We consider an extreme case of zero ionic permeability, $\tilde{D}_\nu(|y|<\ell) = 0$, and negligible electric field within the interface. In addition, the temperature in the bulk is assumed constant, $\theta_0$. For this scenario, one solves the Poisson-Boltzmann equations (Eq.~(\ref{eq:simpGauss})) separately for the bulk and interface. The appropriate boundary conditions are continuity of the electric potential and satisfiying Gauss' law at the two boundaries. 
\begin{equation}\label{eq:boundaryCond}
\begin{aligned}
\tilde{\phi}(-\ell_-) &= \tilde{\phi}(-\ell_+), \\ 
\varepsilon_r^b\partial_{\tilde{y}}\tilde{\phi} |_{-\ell_-} &= \varepsilon_r^i\partial_{\tilde{y}}\tilde{\phi} |_{-\ell_+} - \frac{\chi}{\alpha}\tilde{\sigma}_-, \\
\tilde{\phi}(\ell_-) &= \tilde{\phi}(\ell_+), \\
\varepsilon_r^i\partial_{\tilde{y}}\tilde{\phi} |_{\ell_-} &= \varepsilon_r^b\partial_{\tilde{y}}\tilde{\phi} |_{\ell_+} - \frac{\chi}{\alpha}\tilde{\sigma}_+.
\end{aligned}
\end{equation}
In addition, and without loss of generality, we set the electric potential to zero at $y\rightarrow -\infty$
\begin{equation}
\tilde{\phi}(y\rightarrow -\infty) = 0.
\end{equation}
For the idealized planar geometry, the solution of the ionic concentration is
\begin{equation}
\tilde{c}_{\nu} = \begin{cases}
    \tilde{c}_{\nu,-\infty}\left(\frac{1+\gamma_- e^{(y+\ell)/\lambda_-}}{1-\gamma_- e^{(y+\ell)/\lambda_-}}\right)^2, & \text{$y<-\ell$}.\\
    0 & \text{$-\ell < y < \ell$}. \\
    \tilde{c}_{\nu,+\infty}\left(\frac{1+\gamma_+ e^{-(y-\ell)/\lambda_+}}{1-\gamma_+ e^{-(y-\ell)/\lambda_+}}\right)^2, & \text{$\ell<y$}.
  \end{cases}
\end{equation}
and the electric potential field is
\begin{equation}
\tilde{\phi} = \begin{cases}
    - 2\tilde{\theta}_0\ln\left(\frac{1+\gamma_- e^{(y+\ell)/\lambda_-}}{1-\gamma_- e^{(y+\ell)/\lambda_-}}\right), & \text{$y<-\ell$}.\\
    - 2\tilde{\theta}_0\ln\left(\frac{1+\gamma_-}{1-\gamma_-}\right) & \text{$-\ell < y < \ell$}. \\
    - 2\tilde{\theta}_0\ln\left(\frac{1+\gamma_+ e^{-(y-\ell)/\lambda_+}}{1-\gamma_+ e^{-(y-\ell)/\lambda_+}}\right) + \tilde{\phi}_m, & \text{$\ell<y$}.
  \end{cases}
\end{equation}
To satisfy the boundary conditions (Eq.~(\ref{eq:boundaryCond})), the constants $\gamma_\pm$ and $\lambda_\pm$ are
\begin{equation}
\begin{aligned}
\lambda_\pm & = \left(\frac{\varepsilon_r^b\tilde{\theta}_0}{\alpha\sum_\nu \tilde{c}_{\nu,\pm\infty}}\right)^{1/2}, \\
\gamma_\pm & = \sqrt{1+\left(\frac{b_\pm}{\lambda_\pm}\right)^2} - \frac{b_\pm}{\lambda_\pm}, \\
b_\pm & = -\frac{2\tilde{\theta}_0\varepsilon_r^b\alpha}{\chi\tilde{\sigma_\pm}}.
\end{aligned}
\end{equation}
and the electric potential difference across the interface is
\begin{equation}
\tilde{\phi}_m = 2\tilde{\theta}_0\left[\ln\left(\frac{1+\gamma_+}{1-\gamma_+}\right) - \ln\left(\frac{1+\gamma_-}{1-\gamma_-}\right) \right]
\end{equation}

\end{appendices}

\end{multicols}

\rule{\textwidth}{0.5pt}

\begin{multicols}{2}
{
\footnotesize
\bibliographystyle{unsrt}
\bibliography{ElectricAcousticPulse}
}
\end{multicols}